\documentclass{article}

\usepackage{PRIMEarxiv}

\usepackage{natbib}
\usepackage{algorithm}
\usepackage{algorithmic}
\usepackage{multicol}
\usepackage[export]{adjustbox}
\usepackage{subcaption}
\usepackage{graphicx}
\usepackage{orcidlink}

\title{Promoting Social Behaviour in Reducing Peak Electricity Consumption}

\author{
  Nathan A. Brooks \orcidlink{0000-0002-4233-8304} \\
  Keele University \\
  Staffordshire \\
  United Kingdom \\ \\
  \And
  Simon T. Powers \orcidlink{0000-0003-0092-808X} \\
  Edinburgh Napier University \\
  Edinburgh \\
  United Kingdom \\ \\ \\
  Corresponding author: \texttt{n.a.brooks@keele.ac.uk}
  \And
  James M. Borg \orcidlink{0000-0002-6662-0849} \\
  Aston University \\
  Birmingham \\
  United Kingdom \\
}

\begin{document}
\maketitle

\begin{abstract}
As we transition to renewable energy sources, addressing their inflexibility during peak demand becomes crucial. It is therefore important to reduce the peak load placed on our energy system. For households, this entails spreading high-power appliance usage like dishwashers and washing machines throughout the day. Traditional approaches to spreading out usage have relied on differential pricing set by a centralised utility company, but this has been ineffective. Our previous research investigated a decentralised mechanism where agents receive an initial allocation of time-slots to use their appliances, which they can exchange with others. This was found to be an effective approach to reducing the peak load when we introduced social capital –- the tracking of favours –- to incentivise agents to accept exchanges that do not immediately benefit them. This system encouraged self-interested agents to learn socially beneficial behaviour to earn social capital that they could later use to improve their own performance. In this paper we expand this work by implementing real world household appliance usage data to ensure that our mechanism could adapt to the challenging demand needs of real households. We also demonstrate how smaller and more diverse populations can optimise more effectively than larger community energy systems.
\end{abstract}

% keywords can be removed

\section{Introduction}

In response to anthropogenic climate change, many countries and international organisations have committed to legally binding greenhouse gas emissions targets. The UK and the EU have both recently updated their legislation to include net zero emissions targets in place for 2050 \citep{skidmore2019climate,Sassoli2021regulation}. This requires moving away from using fossil fuels for energy generation and moving towards renewable sources such as photovoltaic cells and wind turbines. Centralised `national grids' are able to `switch on and off' traditional fossil fuel power plants in order to increase or decrease the energy supply to meet the demand of the users. As the proportion of energy being generated from renewable sources increases this raises a problem -- how can load-balancing (the matching of supply and demand) be managed when the output is inherently dependent on weather conditions. This load-balancing problem is easier to address on a small scale, and as such governments and energy providers are supporting the development of `Community energy systems', where local communities such as a small town own and manage their own renewable energy resources \citep{walker2008community, gruber2021current}.

Decentralised community energy systems allow for a higher share of renewable technologies to be integrated into energy generation \citep{chiradeja2004approach}; minimise transmission losses between the source of energy generation and the end users \citep{pepermans2005distributed}; and improve energy security as the energy supply is less impacted by geopolitical factors \citep{alanne2006distributed}. As social awareness of environmental issues increases, the willingness of communities to invest in community energy systems is also expected to increase \citep{pasimeni2019community}. While there are clear benefits to widespread adoption, the shift towards community energy systems means that communities now become involved in some of the tasks that were previously handled by a centralised national grid. In particular, they now become involved in the balancing of supply and demand. A key problem here is how to reduce \emph{peak} demand, i.e. the maximal amount of electricity that is demanded at any one moment in time. The traditional approach to reducing peak demand is differential pricing set by a central utility company. Simply put, households are incentivised to run their appliances at times of low demand through lower pricing at these times \citep{stern1986effectiveness,dutta2017literature}. This has traditionally involved utility companies offering cheaper electricity overnight. Could a community energy system use the same mechanism for reducing peak demand? Photovoltaic cells would require load to be spread out over daylight hours in order to reduce the need for costly energy storage. Then there is the question of how prices should be set and who should set them? People are unlikely to take part in such a scheme unless they perceive that they are being treated fairly. If a community's peak demand is too high then it is unlikely that it will be able to be met by the community's renewable energy sources and so the community is likely to have to resort to buying in electricity from non-renewable sources. Alternatively, if the demand could be spread out more evenly throughout the day then all of it may be met from their renewable sources.

To address these issues we consider an alternative mechanism for load balancing in a community energy system, which focuses on encouraging users to handle load-balancing by actively altering their behavior. We assume that each household has an agent program running on their smart meter, into which they can input their preferred time-slots for when they would like to run high-powered but time-flexible appliances such as washing machines, dishwashers and tumble dryers. The aim is then to allocate actual time-slots to each household agent for when they run their appliances. With this approach a community could collectively own a source of renewable energy, reducing the initial cost, and then use high-powered appliances when the energy is available, as opposed to sharing energy that they generated individually \citep{soto2021peer}. On the one hand this is a classic multi-objective optimisation problem of reducing peak consumption (the maximum amount of energy demanded in a time-slot) while satisfying each households' preferences as far as possible. On the other hand issues of fairness are central. If households are to be motivated to use the mechanism then they will need to perceive the resulting allocation of time-slots to households as being fair (distributive justice \citep{rescher1966distributive}). It is therefore important that agents are able to maximise their collective performance, but also all individuals should be able to perform well individually. Furthermore, households will need to be able to understand why some of their time-slot preferences have not been met and why the preferences of some other households may have been met instead. In other words, they need to perceive the allocation procedure as treating them fairly (procedural justice \citep{hollander2008procedural}). Many agent-based resource allocation approaches, like our mechanism, aim to function as socio-technical systems. This means users can potentially take control of the agents representing them. Achieving this requires agents' actions and decision-making processes to be easily explainable and interpretable by typical users, promoting procedural justice \citep{andras2018trusting, bellman2017socially}.

Beyond the community energy system application our mechanism is addressing a classic resource allocation problem and as such we must consider the balancing of maximising the total utility gained from the resource and the equity of resource allocations. Many approaches to resource allocation, such as the use of virtual auctions, have been found to maximise the utility that can be gained from a resource, and can be adapted to also increase the equity of distribution \citep{buyya2002economic, pla2015multi}. These approaches are, however, centered around trading some form of currency, either real or virtual credits \citep{zhang2012auction}. In the case of a decentralised community energy system, even if all households are allocated an equal amount of virtual credits, removing the ability for wealthier individuals to have more control over the resource, this approach would still be dependent on competition between agents. Investment and use of a community energy system is reliant on cooperation from a community \citep{pasimeni2019community} and as such we should consider whether approaches that are based on cooperation instead of competition can also be effective allowing for stronger social relations if the system is used in a socio-technical manner \citep{bellman2017socially}.

To maximise the efficiency of decentralised community energy systems, many existing agent based approaches focus on energy sharing between buildings based around financial incentives such as differential pricing \citep{stanczak2015dynamic, zhong2014ides, liu2020optimization, akasiadis2017cooperative}, similar to the traditional approach used by utility companies. However, even when these approaches are based on the utilitarian principle of maximising the aggregate utility, this can often lead to inequitable allocations for those already socially disadvantaged \citep{kim2019social, simmons2008tou}. Financial incentives such as differential pricing or fines for utilising more resources than allocated or at an inefficient time can also reduce intrinsic motivation by allowing users to act effectively feel guilt free if the financial cost is not large enough \citep{promberger2013financial, burson2019mo}. As fuel poverty increases there is an increasing need to consider the concept of \emph{energy justice}, which means ensuring that we move towards an energy system that considers both the traditional economic needs of societies but also the environmental issues surrounding climate change and social justice considerations for end users, such as reducing energy poverty \citep{heffron2017concept, jenkins2016energy, sovacool2015energy}. One approach is to instead allow households to borrow energy from neighbors and repay that energy when they are able without a direct financial transaction \citep{prasad2019multi}. However, as energy generation will rarely match demand when using renewable energy sources, this approach requires households to be equipped with expensive battery storage and their own individual sources of energy generation, making implementation costly for the individual. This cost not only makes communities as a whole less likely to adopt the approach \citep{pasimeni2019community}, but further risks disadvantaging those who cannot afford the initial cost of household renewable energy generation such as photovoltaic cells and battery storage \citep{kim2019social}. Our approach does not require each household to generate energy individually and so avoids this issue.

The research introduced in this paper builds upon prior studies by incorporating real-world demand data \citep{petruzzi2013self, petruzzi2014experiments, brooks2020mechanism}. This incorporation aims to investigate the influence of actual demand peaks on the decentralized system's ability to efficiently manage load balancing through agent interactions. The study further examines the response of various simulated community energy systems, varying in size and demographic composition, to these real-world demand peaks. By delving into these dynamics, the research enhances comprehension of the prerequisites necessary for the successful deployment of decentralized community energy systems.

\section{Energy Exchange Simulation}

Petruzzi \citeyearpar{petruzzi2013self} proposes a mechanism for agents trading time-slots for access to a shared energy resource, inspired by the building of social capital between agents (households, or software agents representing them). In their mechanism agents are initially allocated time-slots at random, but can then propose exchanges of time-slots with other agents. Agents have two possible strategies. Selfish agents only accept exchanges that provide them with one of their preferred time-slots. Social agents, on the other hand, accept not only these beneficial exchanges, but also accept an exchange request if they owe a favour to another agent (provided the exchange will not cause them to lose one of their preferred time-slots). An agent owes a favour to another if the other agent has previously accepted an exchange request from them. Petruzzi showed that under this mechanism, a group where every agent was social had on average more of their time-slot preferences satisfied than a group where every agent was selfish. They construed the recording of favours given and received, and the acting upon this by social agents, as the accumulation of a form of electronic social capital \citep{putnam1994social,petruzzi2014experiments}. The concept of social capital in the form of favours is simple to understand, which is crucial in ensuring the system is seen as procedurally transparent and fair, so that users understand how the system works and can see any biases.

\begin{algorithm}[ht!]
\begin{scriptsize}
\begin{multicols}{2}
    \begin{algorithmic}[1]
        \STATE $s \leftarrow $ current simulation
        \STATE $d \leftarrow $ current day
        \STATE $A \leftarrow $ set of $a$ agents
        \STATE $f \leftarrow $ single population counter = $0$
        \WHILE{$s$.is\_complete() == false}
            \FOR {each $a \in  A$ }
                \STATE $a$.receive\_random\_allocation()
            \ENDFOR
            \STATE $e \leftarrow $ inactive exchange count = $0$
            \newline
            \WHILE{$d$.is\_complete() == false}
                \STATE $V \leftarrow $ set of $v$ adverts
                \FOR {each $a \in  A$ }
                    \STATE $v \leftarrow a$.select\_unwanted\_time\_slots()
                    \STATE $V$.list\_advert($v$)
                \ENDFOR
                \newline
                \FOR {each $a \in  A$ }
                    \IF {$a$.received\_request() == true }
                        \STATE go to next agent
                    \ENDIF
                    \IF {$a$.satisfaction() == $1$ }
                        \STATE go to next agent
                    \ELSE
                        \STATE $r \leftarrow a$.identify\_exchange($V$)
                        \STATE $a$.request\_exchange($r$)
                    \ENDIF
                \ENDFOR
                \newline
                \FOR {each $a \in  A$ }
                    \IF {$a$.received\_request() == true }
                        \STATE $a$.accept\_exchange\_if\_approved()
                    \ENDIF
                \ENDFOR
                \newline
                \FOR {each $a \in A$ }
                    \IF {$a$.made\_exchange() \AND $a$.agent\_type == Social}
                        \STATE $a$.update\_social\_capital()
                    \ENDIF
                \ENDFOR
                \newline
                \STATE $d$.no\_trades() == true
                \FOR {each $a \in A$ }
                    \IF {$a$.made\_exchange() == true}
                        \STATE $d$.no\_trades() == false
                    \ENDIF
                \ENDFOR
                \IF {$d$.no\_trades() == false}
                    \STATE $e$ = $e$ + $1$
                \ELSE
                    \STATE $e$ = $0$
                \ENDIF
                \IF {$e$ == $10$}
                    \STATE $d$.is\_complete() = true
                \ENDIF
            \ENDWHILE
            \newline
            \FOR {each $a \in  A$ }
                \STATE {$a_2 \leftarrow $ random agent to observe}
                \IF {$a$.satisfaction() \textless \ $a_2$.satisfaction() }
                    \STATE {$x \leftarrow $ random value between $0$ and $1$}
                    \IF{$a$.learning\_probability($a$.satisfaction()) \textgreater \ $x$}
                        \STATE $a$.copy\_strategy($a_2$)
                    \ENDIF
                \ENDIF
            \ENDFOR
            \newline
            \IF {$s$.single\_type\_remaining() = true}
                \STATE $f$ = $f$ + $1$
            \ENDIF
            \IF {$f$ == $100$}
                \STATE $s$.is\_complete() = true
            \ENDIF
            \STATE $d$ = $d$ + $1$
        \ENDWHILE
    \end{algorithmic}
    \end{multicols}
    \end{scriptsize}
    \caption{The Energy Exchange Simulation.}
    \label{alg:algorithm}
\end{algorithm}

The Energy Exchange Simulation has been developed in order to better understand how social capital, in the form of agents owing each other favours, can influence direct interactions between agents in pairwise situations, and how this in turn can impact on the success of the population of households in solving the multi-objective optimisation problem of load balancing. More specifically, it allows us to explore the conditions that allow for members of a community energy system to distribute energy from a shared resource in a way that satisfies the members demands without requiring individual household generation, financial transactions or battery storage. The experimentation discussed in this paper builds upon the research by Petruzzi \citeyearpar{petruzzi2013self}. Their work demonstrated how agents can effectively utilize social capital by keeping track of owed and received favors. This strategy enables the agents to dynamically allocate time slots for appliance usage, thereby enhancing load balancing within a decentralized system. Following on from the research by Petruzzi \citeyearpar{petruzzi2013self}, the model was built to represent a smart energy network consisting of $96$ individual agents. Each day agents request four hour-long time-slots in which they require electricity. We consider all requests to be for an equal amount of energy. These values were initially selected by Petruzzi to allow for all time-slots to be allocated with an availability of 16 kWh units of energy for each hour of the day. Unless otherwise specified, parameters were kept consistent with the previous research by Petruzzi \citeyearpar{petruzzi2013self} to allow for a comparison of results. We have maintained these values such as the $96$ population size for our initial results, but allow the total availability of energy to linearly scale with the total demand so that we can adjust the population size of the community energy system (number of households) as required (this assumption corresponds to a community energy system with more members installing proportionately more renewable energy sources). Time-slots are initially allocated to household agents at random at the start of the day. This randomness allows for privacy, and is procedurally fair as there are no biases, however agents are highly unlikely to have this random allocation match all of their requested time-slots. To address this problem, after the initial allocation agents can partake in pairwise exchanges where one agent requests to swap one of its time-slots with a second agent, and the second agent decides whether or not to fulfil the request. We define an agent's \emph{satisfaction} as the proportion of its time-slot preferences that have been satisfied, and track the mean value of this as a measure of how well the mechanism is satisfying the agents' preferences.

Agents can follow either a social or a selfish strategy which impacts how they react to incoming requests for exchanges. Selfish agents will only accept exchanges that are in their immediate interest. This means that selfish agents need to be offered a time-slot that they have initially requested and do not already have in order for them to agree to the exchange. Social agents also agree to these mutually beneficial exchanges. In addition, social agents also make decisions based on social capital, in the form of repaying previous favours given to them by other agents. Specifically, when a social agent's request is accepted they record it as a favour given to them. When a social agent receives a request from another agent who previously gave them a favour they will accept the request, if it is not detrimental to their own satisfaction, and then record that the favour has been repaid. This leads to a system of social agents earning and repaying favours among one another increasing the number of accepted exchange requests.

Exchanges begin every day once each agent has received their initial allocation and decided which of these time-slots they wish to keep. They then anonymously advertise slots that they have been allocated but do not want to an `advertising board'. Several exchange rounds then take place during the day with exchanges continuing until no request has been accepted in the past $10$ rounds of exchanges and the average satisfaction has therefore stopped increasing. In each exchange round agents can request a time-slot from the board so long as they have not already received a request from another agent during that round. Agents accept or refuse requests based on their strategy, as described above. Only social capital, i.e. social agents' memory of favours, remains between days. The simulation continues until the entire population of agents has adopted a single strategy, and then for $100$ further days so that performance can be evaluated once only one strategy has been present for a prolonged period of time.

The Energy Exchange Simulation also incorporates social learning allowing agents to change from selfish to social or vice versa (note that both social and selfish agents undergo `social' learning, which we refer to simply as `learning' from now on to avoid confusion). This works as follows. Each agent observes a randomly selected second agent. If the observed agent has a higher satisfaction than the agent in question, then the first agent has a chance to copy the second agents strategy with the likelihood of copying their strategy determined by the following equation:

\begin{equation}\label{equation}
    P_s =2\left(\frac{1}{1 + exp(-\beta(a_{2s} - a_{1s}))}\right) - 1
\end{equation}

The equation was adapted from the pairwise comparison social learning process given in the third figure in Isakiv and Rand's work on the evolution of coercive institutional punishment \citep{isakov2012evolution}. $P_s$ refers to the probability of an agent changing strategy, where $a1s$ refers to the observing agents satisfaction and $a_{2s}$ refers to the satisfaction of the agent being observed. The $\beta$ value determines the selection pressure and was set to $1$ unless specified otherwise. The results are normalised to give a value between $0$ and $1$. Learning is thus payoff-biased \citep{boyd1988culture}, with strategies giving higher individual satisfaction more likely to spread in the population. Agents that move from a social strategy to the selfish strategy retain their accumulated social capital. Pseudo-code for the simulation procedure is given in Algorithm~\ref{alg:algorithm}.

\section{Experimentation}
\subsection{Logical Comparison of Strategies}
Prior to any experimentation, we can logically deduce that agents choosing the social strategy should increase the average satisfaction of the agents as a collective more so than agents choosing the selfish strategy. This is because given that no agent gives away a slot that it has requested, and agents only request time-slots that they want, any exchange must increase the satisfaction of the agent making the request, and either have no effect or increase the satisfaction of the agent receiving the request. With exchanges only having a positive or neutral impact on the agents involved, more exchanges must lead to a greater mean satisfaction. In pure populations, where all agents are utilising the same strategy, strategies that allow for more exchanges will have a higher mean satisfaction. As agents using the social strategy will accept any exchange that a selfish agent would accept but will also make exchanges based social capital, a population of purely social agents will be more satisfied than a population of purely selfish agents.

In order to confirm this behavior we ran the simulation with pure populations and compared the mean satisfaction of the agents. The results presented herein are derived from the average outcomes of $100$ simulation runs. This approach was adopted to ensure the stability and uniformity of the average outcomes upon retesting, effectively mitigating the influence of the inherent variability in simulation outputs on the presented data. After a single day, purely selfish populations finished with a mean satisfaction of $0.683$, where a satisfaction of $1$ would mean that all agents had the time slots they requested. Purely social populations performed similarly after a single day with a mean satisfaction of $0.696$. This is because after a single day very little social capital will have been accumulated by the agents, and so the number of exchanges will be similar to the selfish populations. After $100$ days, once social capital has been built up by the social agents, where selfish agents see no notable change in performance social agents mean satisfaction jumps to $0.834$. This was found to be a statistically significant result with a Mann-Whitney U test  $(p < 0.01)$, with the highest performing selfish population having a lower average satisfaction than the worst performing social population. This confirms our hypothesis that purely social populations will be able to outperform purely selfish populations. We can also calculate a theoretical optimum average satisfaction by comparing all requested time-slots with all those available within the simulation. This optimum value is a purely theoretical comparison of the two sets of values, the requested time-slots and the available time-slots, at the start of the simulation and does not account for issues such as fairness and trust. Any system in which the performance nears the theoretical optimum while accounting for these issues can thus be considered high performing. Social populations are able to perform very close to the theoretical optimum performance, which gave a theoretical optimum mean satisfaction of $0.852$ averaged across the same $100$ runs of the simulation.

\subsection{Expansion on Previous Research}
Our preliminary investigations showed how a social strategy based around social capital, as opposed to social agents always accepting neutral exchanges regardless of their history with the agent making the request, is essential in allowing social agents to outperform selfish agents in mixed populations \citep{brooks2020mechanism}. With payoff-biased social learning we also demonstrated how self-interested agents will adopt the social strategy with increasing frequency later in the simulation, as the build up of social capital widens the difference in mean satisfaction between the selfish and social agents. This research demonstrated that social capital can be an effective mechanism for promoting social behaviour and maximising agent satisfaction. However, prior work only considered this in a population with an artificially flat demand curve, where every hour of the day had an equal amount of agents requesting to use their appliances\citep{brooks2020mechanism}. This is clearly not a realistic expectation for the average energy demand within a community. Many households follow typical behaviours such as leaving their house for work during the day, and cooking and using electrical appliances in the evenings. In order to account for this we have used data from the Household Electricity Survey (HES) to integrate realistic energy demand into our model \citep{zimmermann2012household}. By having agents demand curves be based off this real data we were able to explore how the mechanism fared in challenging real world scenarios.

The Household Electricity Survey was conducted between 2010 and 2011, and was the most detailed monitoring of electricity use ever carried out in the UK, providing real demand curves for specific household appliances from $250$ UK households \citep{palmer2013further}. The model provides dis-aggregated data about specific types of appliances, and can be sorted by key metrics such as the size of the household, the number of residents, and the residents ages. This is particularly important as regardless of the environmental impact, it is unlikely that many households would be comfortable adjusting their daily routine such as mealtimes in order to flatten the average energy demand curve of the community. Despite this, modern smart appliances such as washing machines, dishwashers and tumble dryers, can be scheduled to run at a specific time and can be considered to be ``switchable" appliances as their usage is inherently less time sensitive \citep{zimmermann2012household}. With this in mind, we re-ran the simulation with the probability of agents requesting a specific time-slot being based off of the real demand data for switchable appliances from the HES.

\subsection{Flat Demand Curve}
In order to understand how using real world demand data impacts the behavior of the system, we ran the simulation first with a flat demand curve, where agents had an equal likelihood of requesting any hour of the day when choosing their four time-slots to request. Unlike our previous research, this demand curve only ensured that there was an equal likelihood of agents requesting each time-slot, it did not guarantee that each time-slot would have an equal number of requests. Running the simulation $100$ times, all $100$ runs finished with the entire population adopting the social strategy from a starting point of half the agents using the social strategy and half using the selfish strategy. The social takeover took an average of $165$ days with the mean satisfaction of the agents being $0.834$ on the day that the selfish strategy was eliminated, and $0.843$ when the simulation finished, and all agents had been able to build up social capital through using the social strategy $100$ days later.

\subsection{Real Demand Curve}
\begin{figure}[ht]
    \centering
    \includegraphics[width = 0.55\textwidth]{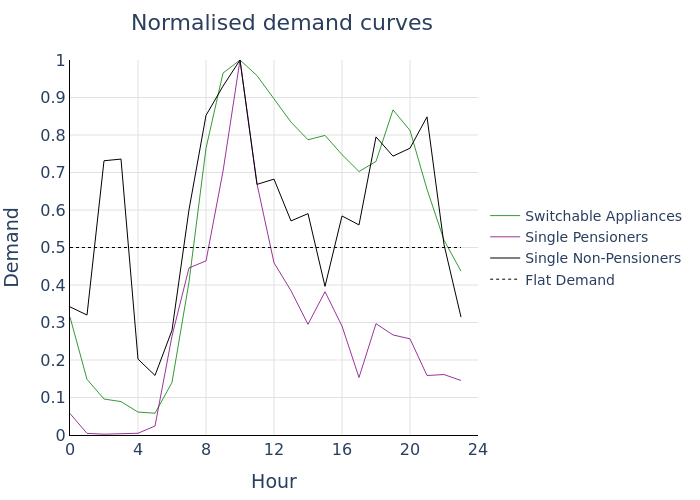}
    \caption{Demand curves used by agents to select hourly time-slots. Demand levels are normalized for easy comparison. "Switchable appliances" refers to modern smart appliances like washing machines, dishwashers, and tumble dryers that can be scheduled. "Single pensioners" and "Single non-pensioners" represent demographics of those living alone. These curves are derived from real appliance usage data in the Household Energy Survey, adapted for the hourly time-slot model \citep{zimmermann2012household}.}
    \label{fig:curves}
\end{figure}

We next changed the demand curve that the agents used to decide which time-slots to request each day using data from the HES. Using real data for the use of ``switchable" appliances, agents were much more likely to have similar demands for time-slots during times of real-world peak demand. The demand curves used in this paper can be seen in figure \ref{fig:curves}. With this change, social agents were only able to take over the population in $56$ of the $100$ runs of the simulation, with the other $44$ runs resulting in a purely selfish population. In runs where the population fully adopted the social strategy, the average time for the social strategy to takeover was $310$ days, with a mean satisfaction of $0.707$ when the strategy was fully adopted, and $0.716$ when the simulation ended $100$ days later. The calculated optimum satisfaction averaged at $0.721$ for all the days in all simulation runs, and so it is clear that when the social strategy does fully takeover the population then the agents are able to perform at a near optimum level. In the $44$ runs where the selfish strategy took over the population the average satisfaction of the agents was $0.578$ at the end of the simulation, a clear drop in performance compared to the social strategy, which was found to be statistically significant with a Mann-Whitney U test $(p<0.01)$, with no overlap between the two sets of results.

\begin{figure*}[!t]
\centering
    \includegraphics[width = 1.0\textwidth]{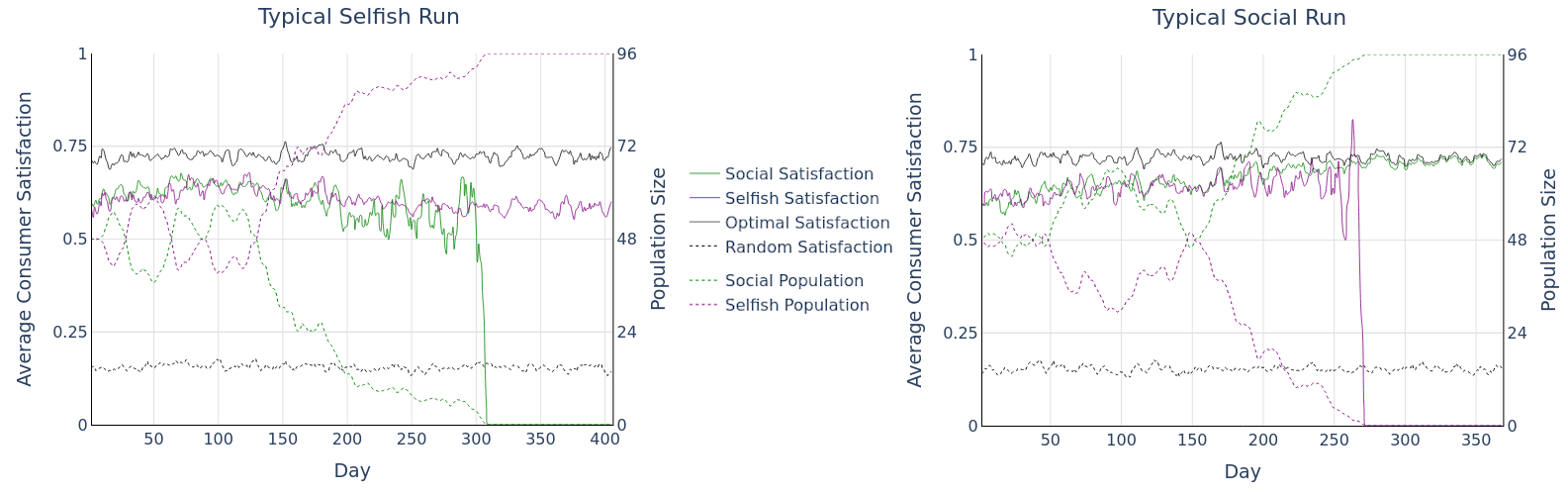}
    \caption{Representative runs of the simulation that resulted in the population becoming either fully social or selfish. The representative runs are chosen as the median run when all runs that became either fully social or fully selfish are ordered by the time it took to reach that state. The values for each day are calculated as a five point moving average.}
    \label{fig:graphs}
\end{figure*}

Figure \ref{fig:graphs} shows representative runs of the simulation that resulted in either a fully social or fully selfish population. The specific runs shown represent the median run when the runs that resulted in a specific strategy are ordered by how many days it took for the population to adopt that strategy. The average satisfaction values shown on the graphs represent a five day moving average in order to minimise the effect of day to day variance and represent all agents using the strategy that eventually takes over at that time. In both the run that resulted in a fully social population, and the run that resulted in a fully selfish population, the average satisfaction was very similar for the first $100$ days, with selfish agents performing very similarly to their social counterparts. Later in the simulation however, the selfish runs average satisfaction dropped. This is because there were no longer enough social agents in the population to increase the average satisfaction by performing more exchanges.

\begin{table*}[!t]
    \centering
    \resizebox{1.0\textwidth}{!}{
    \begin{tabular}{|c||*{8}{c|}}
    \hline{Day}
    &\makebox{1}&\makebox{50}&\makebox{100}&\makebox{150}&\makebox{200}&\makebox{250}&\makebox{300}&\makebox{350}\\\hline
    Social Population &48 &52 &66 &47 &77 &92 &96 &96\\\hline
    Average Social Satisfaction &0.641 &0.654 &0.700 &0.654 &0.714 &0.687 &0.716 &0.734\\\hline
    Average Social SD &0.244 &0.215 &0.236 &0.239 &0.257 &0.281 &0.257 &0.262\\\hline
    Average Selfish Satisfaction &0.526 &0.590 &0.742 &0.689 &0.711 &0.4 &0 &0\\\hline
    Average Selfish SD &0.193 &0.233 &0.199 &0.217 &0.233 &0.122 &0 &0\\\hline
    \end{tabular}
    }
    \caption{Statistics for specific days from the typical social run displayed in Figure~\ref{fig:graphs}. Social Population shows the number of the 96 agents that are using the social strategy. SD refers to standard deviation from the mean satisfaction.}
    \label{tab:typicalSocial}
\end{table*}

\begin{table*}[!t]
    \centering
    \resizebox{1.0\textwidth}{!}{
    \begin{tabular}{|c||*{8}{c|}}
    \hline{Day}
    &\makebox{1}&\makebox{50}&\makebox{100}&\makebox{150}&\makebox{200}&\makebox{250}&\makebox{300}&\makebox{350}\\\hline
    Selfish Population &48 &60 &40 &65 &84 &88 &92 &96\\\hline
    Average Selfish Satisfaction &0.526 &0.629 &0.663 &0.6 &0.628 &0.545 &0.568 &0.596\\\hline
    Average Selfish SD &0.246 &0.216 &0.241 &0.227 &0.227 &0.243 &0.236 &0.226\\\hline
    Average Social Satisfaction &0.589 &0.590 &0.665 &0.710 &0.583 &0.563 &0.689 &0\\\hline
    Average Social SD &0.213 &0.237 &0.223 &0.212 &0.236 &0.207 &0.108 &0\\\hline
    \end{tabular}
    }
    \caption{Statistics for specific days from the typical selfish run displayed in Figure~\ref{fig:graphs}. Selfish Population shows the number of the 96 agents that are using the Selfish strategy. SD refers to standard deviation from the mean satisfaction.}
    \label{tab:typicalSelfish}
\end{table*}

Tables \ref{tab:typicalSocial} and \ref{tab:typicalSelfish} show specific days in the typical simulation runs displayed in Figure~\ref{fig:graphs}. Upon comparing the outcomes after a single day within the simulation, it becomes evident that the two runs exhibit distinct disparities in average social satisfaction. This observation underscores the considerable variation in the distribution of time slots between simulation runs, which can be attributed to the initial random assignment of these time slots. It is also clear that the strategy used by agents fluctuates greatly throughout a typical simulation, with neither run having a steady trend in strategy used towards its final state. This fluctuation greatly benefits the selfish strategy, as agents can accrue social capital while being social before switching to the selfish strategy and benefiting from their existing social capital with social agents, while no longer assisting others by accepting request based on social capital. These tables also demonstrate how on a given day their is often little difference between the mean performance of the two strategies, with the strategy performing superior often fluctuating day to day. This promotes the fluctuation between strategies, benefiting the selfish strategy as discussed. As we have previously shown, the population fully adopting the social strategy is in the best interest of all agents involved as it increases the mean satisfaction of the agents. These results show that in order for population to consistently adopt the social strategy, there needs to be enough of a consistent performance difference between the two strategies that agents are both more likely to adopt the social strategy, and unlikely to switch when observing a selfish agent.

\subsection{Effect of Population Size}
\begin{table*}[!t]
    \centering
    \resizebox{1.0\textwidth}{!}{
    \begin{tabular}{|c||*{8}{c|}}
    \hline{Population Size}
    &\makebox{24}&\makebox{48}&\makebox{72}&\makebox{96}&\makebox{120}&\makebox{144}&\makebox{168}&\makebox{192}\\\hline
    \hline
    Social Takeovers &84 &87 &82 &56 &31 & 11 & 5 &1\\\hline
    Average Takeover Days (social) &111 &184 &229 &310 &318 &367 &477 &380\\\hline
    Average Takeover Satisfaction (social) &0.598 &0.663 & 0.694 &0.707 &0.722 &0.736 & 0.725 &0.731\\\hline
    Average Satisfaction $100$ days after takeover(social) &0.623 &0.688 &0.704 &0.716 &0.727 &0.733 &0.738 &0.730\\\hline
    \hline
    Selfish Takeovers &16 &13 &18 &44 &69 &89 &95 &99\\\hline
    Average Takeover Days (selfish) &131 &219 &277 &365 &355 &334 &317 &302\\\hline
    Average Takeover Satisfaction (selfish) &0.421 &0.514 &0.555 &0.584 &0.604 &0.617 &0.625 &0.632\\\hline
    \end{tabular}
    }
    \caption{The performance difference between different population sizes.}
    \label{tab:popSizeComparison}
\end{table*}

In order to get a better understanding of how this system could perform in real world scenarios we set out to explore how changing the initial population could influence the systems performance. The first parameter explored was the size of the population. This is important to understand as in a real world smart-city it may be practical to divide the population into multiple smaller community energy systems. While indirect reciprocity facilitated by reputation, which we model as social capital, has been found to be beneficial to the performance of populations in public goods games, this has rarely been the case with larger populations \citep{nowak1998evolution}. The network topology used has a major impact on the viability of social behaviour, and heavily clustered smaller societies have been found to promote more cooperation than larger networks \citep{santos2008social, neumann2020indirect}. For our mechanism it is therefore important to determine which population sizes facilitate social behavior and thus maximise the average agent satisfaction. To compare with the $96$ population simulation runs already discussed, we re-ran the simulation with a variety of population sizes between $24$ and $192$ agents. These numbers were chosen in order to be between $25\%$ and $200\%$ of the initial population of $96$ agents. In order to minimise the variables in the system the availability of energy in the system was linearly scaled to meet the demand required. Table~\ref{tab:popSizeComparison} shows how the systems behaviour changed by adjusting only the population size. The smallest population of $24$ agents adopted the social strategy far more frequently than the larger populations, with the social strategy taking control on $84$ of the $100$ simulation runs. This is a clear contrast with the largest population of $192$ agents, which adopted the social strategy only once in all $100$ simulation runs. However, while the smallest population adopted the social strategy more often, the average satisfaction at the end of the simulation was only $0.623$, which is similar to the average satisfaction seen with the largest populations selfish runs which averaged $0.632$.

This result shows a clear trade-off that must be made when deciding on the population size for a community energy system using this energy exchange mechanism. Smaller populations are more likely to adopt the social strategy and are therefore able to perform at a near optimal level. Larger populations are less likely to adopt the social strategy, but their potential performance is higher when they are able to do so. The reason we see this behavior is because in smaller populations, each individual is more likely to interact with each other individual and so social capital can be built up much more quickly. At the end of the simulation runs, with the smallest population size agents had an average of $32.55$ unspent social capital where as in the largest population the average was only $21.44$ unspent social capital. An alternative approach for larger populations could be to use a global social capital mechanism, however this could remove the user privacy that comes from agents tracking social capital in a pairwise manner and storing it themselves. In larger populations social capital is slower to build but with more agents there are more time-slots in the system, and so more agents to potentially trade with. It is also worth noting that there was little difference in the number of populations taken over by the social strategy between the simulations with populations of $24$, $48$ and $72$ agents, which were taken over by the social strategy $84$, $87$ and $82$ times respectively. This shows how decreasing the populations size only benefits the social strategy up to a certain point, and so it is important not to reduce the population size beyond this in order to maximise the average satisfaction, which was higher with larger social populations.

\subsection{Effect of Diverse Demographics}
We next explored how having mixed demographics within the population influenced the performance. In the real-world different demographics can have very different usage patterns for their appliances. A large family will need to use their washing machine more frequently and at different times of day than a young individual living alone. From the Household Electricity Survey we took the switchable appliance usage data for single pensioners and single non-pensioners and generated two separate demand curves. These two groups were found to have clear differences in their usage patterns, with pensioners having a large spike in usage between 10:00 and 11:00 with usage steadily declining until midnight, and non-pensioners having comparatively more consistent usage throughout the day, as a larger number of single non-pensioners choose to use their appliances at night. These two groups have been selected for discussion as they were the most clearly diverse groups and as such their pairing had the largest increase in agent satisfaction. These two demand curves are shown in figure \ref{fig:curves}. We ran the simulation such that half the agents would use the demand curve of the pensioners to influence their requests for time-slots and the other half used the demand curve of the non-pensioners.

In this mixed demographic scenario the social strategy was able to take over the population in $90$ of the $100$ simulation runs. This took an average of $272$ days and resulted in a mean satisfaction of $0.733$ immediately after all agents became social, and a mean satisfaction of $0.750$ when the simulation ended $100$ days later. For the $10$ simulation runs that adopted the selfish strategy, it took an average of $385$ days for the selfish strategy to takeover and the average satisfaction was $0.600$ at the end of the simulation. The lowest average satisfaction from a social takeover at the end of a run was higher than the highest average satisfaction from a selfish run, with the two sets difference between the two types of runs being clearly statistically significant with a Mann-Whitney U test $(p < 0.01)$. By diversifying the population such that they have varied demand curves, the social strategy was able to takeover more consistently and the mean satisfaction increased in all scenarios compared to when all agents used a single demand curve. A demographically diverse population is clearly a very effective way to allow for this mechanism to perform as optimally as possible.

\begin{table*}[!t]
    \centering
    \resizebox{1.0\textwidth}{!}{
    \begin{tabular}{|c||*{8}{c|}}
    \hline{Population Size}
    &\makebox{24}&\makebox{48}&\makebox{72}&\makebox{96}&\makebox{120}&\makebox{144}&\makebox{168}&\makebox{192}\\\hline
    Social Takeovers &85 &94 &93 &90 &81 &80 &57 &42\\\hline
    Average Takeover Days (social) &107 &169 &192 &272 &321 &382 &409 &463\\\hline
    Average Takeover Satisfaction (social) &0.609 &0.683 &0.712 &0.733 &0.747 &0.756 &0.758 &0.770\\\hline
    Average Satisfaction $100$ days after takeover(social) &0.656 &0.706 &0.731 &0.750 &0.754 &0.758 &0.768 &0.767\\\hline
    \hline
    Selfish Takeovers &15 &6 &7 &10 &19 &20 &43 &58\\\hline
    Average Takeover Days (selfish) &108 &174 &282 &385 &340 &549 &552 &533\\\hline
    Average Takeover Satisfaction (selfish) &0.413 &0.505 &0.574 &0.604 &0.617 &0.626 &0.648 &0.654\\\hline
    \end{tabular}
    }
    \caption{The performance difference between different population sizes when using two demand curves for single pensioners and single non-pensioners.}
    \label{tab:mixedPopSizeComparison}
\end{table*}

Using the two separate demand curves we next re-ran the simulation with varying population sizes, as with Table~\ref{tab:popSizeComparison}, to see whether alternative population sizes had notably changed results. As seen in Table~\ref{tab:mixedPopSizeComparison}, a mixed population allowed for improved improved mean satisfaction for runs where the social strategy took over, and also allowed the social strategy to take over more consistently compared to when a single demand curve was used as seen in Table~\ref{tab:popSizeComparison}. This impact was most evident with the the larger population sizes, the populations of $96$, $120$ and $144$ agents improved on the number of runs adopting the social strategy from $56$, $31$ and $11$ runs to $90$, $81$ and $80$ runs respectively. This clearly shows how a population with diverse demands can perform well with a much larger range of population sizes than a population with more similar demands. It is also worth noting how the smallest population tested, $24$ agents, adopted the social strategy less frequently than the slightly larger sizes of $48$, $72$ and $96$ agents. This further demonstrates how it is important to select a population size that allows for populations to adopt the social strategy while being large enough for a high average satisfaction, but goes beyond our previous understanding by suggesting that having too small of a population size can reduce the social strategies ability to take over consistently. This is because while smaller populations build up social capital more easily, with less potential trade partners there is a greater risk of only a small number of trades happening in a day, and those trades being between selfish agents.

\subsection{Effect of Selection Pressure}
\begin{table*}[!t]
    \centering
    \resizebox{1.0\textwidth}{!}{
    \begin{tabular}{|c||*{7}{c|}}
    \hline{$\beta$ Value}
    &\makebox{0.5}&\makebox{0.75}&\makebox{1}&\makebox{2}&\makebox{3}&\makebox{4}&\makebox{5}\\\hline
    \hline
    Social Takeovers &100 &95 &90 &51 &38 &30 &25\\\hline
    Average Takeover Days (social) &325 &282 &272 &219 &176 &152 &111\\\hline
    Average Takeover Satisfaction (social) &0.737 &0.734 &0.733 &0.735 &0.726 &0.730 &0.726\\\hline
    Average Satisfaction $100$ days after takeover(social) &0.741 &0.743 &0.750 &0.749 &0.744 &0.734 &0.742\\\hline
    \hline
    Selfish Takeovers &0 &5 &10 &49 &62 &70 &75\\\hline
    Average Takeover Days (selfish) &- &332 &385 &258 &176 &145 &128\\\hline
    Average Takeover Satisfaction (selfish) &- &0.630 &0.604 &0.599 &0.607 &0.603 &0.606\\\hline
    \end{tabular}
    }
    \caption{The performance difference between different levels of selection pressure when using two demand curves for single pensioners and single non-pensioners.}
    \label{tab:selectionPressureComparison}
\end{table*}

Our final set off experiments involved a population of $96$ agents using the two separate demand curves for single pensioners and single non-pensioners. We altered the selection pressure for social learning between agents by altering the $\beta$ value shown in equation \ref{equation}. With larger $\beta$ values smaller differences in satisfaction are more likely to cause agents to change their strategy. This thus allows us to control how sensitive to satisfaction differences agents are when learning their strategies. Table \ref{tab:selectionPressureComparison} shows how the simulation performs as the $\beta$ value used in our learning equation is altered. The results show how as the selection pressure increases the number of simulation runs in which the social strategy successfully takes over the population decreases. This makes logical sense, as with greater selection pressure agents will change their strategy more frequently which will lead to more selfish agents benefiting from social capital they gained when social, preventing social agents from having an advantage. We also see that with a greater level of selection pressure both strategies are able to take over the population considerably more quickly. Conversely, when the selection pressure is lowered the social strategy is much more effective, as agents are able to build up social capital more effectively when their strategy remains consistent. With a $\beta$ value of $0.5$ the social strategy was able to take over in all $100$ simulation runs. Later experiments even showed that with the largest population size of $192$ agents, the social strategy was able to take over in $97$ of the $100$ simulation runs when a $\beta$ value of $0.5$ was used, however this takeover took $566$ days on average. Lowering the selection pressure is clearly very successful in promoting social behaviour, however it can greatly slow down the time required for agents to settle on a single strategy.

\section{Conclusion}
By incorporating real world demand data into our model, we have shown how realistic demand peaks make it substantially harder for social agents to take over a population consistently when compared to an artificial flat demand curve. We have shown that there is a clear improvement in performance when a population consists of mixed demographics, and so this approach to load balancing can be effective in some realistic scenarios. We have also shown how it is important that the population size is large enough to allow for a high potential average satisfaction but not so large that the social agents are unable to build up social capital. For a real world community energy system, particularly in larger smart cities, this suggests that by organising larger populations into specific community clusters, with a variety of household types, we could see greatly improved demand side load balancing than with a larger monolithic population. Allowing agents to form self-organised clusters working to optimise their collective performance has been shown to be an effective approach to supply and demand matching within large-scale energy systems \citep{vcauvsevic2017dynamic}. Future research will consider how larger population sizes should be clustered such that our energy exchange mechanism can be used for larger smart cities. The process of clustering and the subsequent inter-cluster communication will also play a vital role in enabling the effectiveness of this load balancing strategy within real-world markets. This mechanism opens the possibility for distinct clusters to engage in energy trading when surplus energy is available within one cluster. Moreover, established energy suppliers would have the capacity to augment the energy generation of communities by providing energy to meet their requirements.

A significant limitation of the existing model lies in its binary treatment of agent preferences—where they are either fully satisfied or entirely dissatisfied with a given time slot. In real-world scenarios, users are more likely to retain some level of preference for time slots near their initial choices. As a result, future research will delve into the dynamic impact on the behaviors of both social and selfish agents as the satisfaction they derive from an allocated time slot becomes contingent on its proximity to their initial preference. This advancement will enable the exploration of intricate social dynamics, including instances where agents relinquish a moderately satisfying time slot to benefit others who could attain a higher level of satisfaction. Additionally, this approach will facilitate the examination of scenarios wherein agents may choose to forego immediate gratification in exchange for the possibility of securing a more favorable time slot in subsequent exchanges. We also showed that by lowering the selection pressure used when agents are considering changing their strategy, social agents can build up more social capital and the social strategy takes over the population much more consistently. This selection pressure is however an aspect of our simulation that is not based on the behavior of real users and so future work should empirically investigate how likely real users would be to alter their strategy.

We have demonstrated a decentralised mechanism for household load balancing that is effective at satisfying agents' preferences. The benefits of a decentralised mechanism are that it is inherently scalable as more agents are introduced \citep{petruzzi2013self}, and helps to promote privacy and trust by not requiring households to submit their time-slot preferences to a centralised authority. There are also more complex algorithms that have the potential to be highly effective at managing other aspects of a households energy usage that we did not cover such as heating systems\citep{kolen2017two, dengiz2020decentralized}. Our work differs in that it is inherently human facing. A real world implementation of our system could easily operate in a socio-technical manner in which individuals can take over from the virtual agent representing them, setting their own preferences for time-slots and making decisions on whether or not to accept requested exchanges. In an actual implementation, the exchanges might occur through software agents that operate on smart meters found within households, or on cloud servers where smart meters serve as the intermediary interface \citep{saoud2022hybrid}. This could involve various levels of user engagement with the exchange process. A time-slot-based approach to demand-side management could be integrated by utility companies alongside community energy systems, enhancing efficiency within a hybrid system that remains consistent in its requirements for changes in consumer behavior \citep{cortez2023demand}. Utilising a system based on social capital represented as `favours' would also be easy for the average user to understand, facilitating procedural justice and promoting social behavior within the community. As our approach doesn't require all users to generate their own energy, require expensive battery storage or require financial transactions, it is also an approach to community energy systems that minimises the cost to individual households and so also contributes to improved \emph{energy justice} \citep{heffron2017concept, jenkins2016energy, sovacool2015energy}.

\section{Data Availability Statement}
The demand curves analysed and model used are available in the following GitHub repository: \url{https://github.com/NathanABrooks/ResourceExchangeArena}.
\newline
The original Household Electricity Survey data analysed during the current study is not publicly available but is available from the UK Data Service on request.

\section{Statements and Declarations}
The authors have no relevant financial or non-financial interests to disclose.

\bibliographystyle{apalike}  
\bibliography{main}

\end{document}